\documentclass[10pt]{article}
\usepackage{graphicx}

\newcommand\pubnumber{SLAC-PUB-12493}
\newcommand\pubdate{July, 2007}

\def\SLAC{Stanford Linear Accelerator Center\\
    Stanford University, Stanford, California 94309 USA}
\def\doeack{\footnote{Work supported by the US Department of Energy,
                     contract DE--AC02--76SF00515.}}

\def\Title#1{\begin{center} {\Large #1 } \end{center}}
\def\Author#1{\begin{center}{ \sc #1} \end{center}}
\def\Address#1{\begin{center}{ \it #1} \end{center}}

\newcommand\pubblock{\rightline{\begin{tabular}{l} \pubnumber\\
         \pubdate \end{tabular}}}
\newenvironment{Abstract}{\begin{quotation} \begin{center}
                       ABSTRACT
     \end{center}\bigskip  }{\end{quotation}}
\newenvironment{Presented}{\begin{quotation} 
      \begin{center}}{\end{center} \end{quotation}}

\def\Acknowledgements{\bigskip  \bigskip \begin{center} \begin{large}
             \bf ACKNOWLEDGEMENTS \end{large}\end{center}}
\def\beq{\begin{equation}}
\def\eeq#1{\label{#1}\end{equation}}
\def\eeqn{\end{equation}}
\def\leqn#1{(\ref{#1})}
\def\VEV#1{\left\langle{ #1} \right\rangle}
\def\Eslash{\not{\hbox{\kern-4pt $E$}}}
\def\Pl{{\mbox{\scriptsize Pl}}}
\begin{document}
\begin{titlepage} 
\pubblock

\vfill
\Title{Dark Matter and Particle Physics}
\vfill
\Author{Michael E. Peskin\doeack}
\Address{\SLAC}
\vfill
\begin{Abstract}
Astrophysicists now know that 80\% of the matter in the universe is 
`dark matter', composed of neutral and weakly interacting elementary
particles that are not part of the Standard Model of particle physics.
I will summarize the evidence for dark matter.  I will 
explain why I expect dark matter particles to be produced at the 
CERN LHC.  We will then need to characterize the new weakly interacting
particles and demonstrate that they are the same particles that are found in 
the cosmos.  I will describe how this might be done.
\end{Abstract}
\vfill
\begin{Presented}
to appear in the Special Topics Issue of the\\  Journal of the Physical Society
in Japan (JPSJ)\\ `Frontiers of Elementary Particle Physics:\\
 The Standard Model
and Beyond' 
\end{Presented}
\vfill
\end{titlepage}
\def\thefootnote{\fnsymbol{footnote}}
\setcounter{footnote}{0}
\tableofcontents
\newpage
\setcounter{page}{1}

\section{Introduction}

One of the themes of the history of physics has been the discovery
that the world familiar to
us is only a tiny part of an enormous and multi-faceted universe.  From
Copernicus, we learned that the earth is not the center of the universe, from 
Galileo, that there are other worlds.  More recently, Hubble's 
extragalactic astronomy taught us that our galaxy is a tiny part of an
expanding universe, and the observation of the cosmic microwave background
by Penzias and Wilson revealed an era of cosmology before the formation of 
structure.  Over the past ten years, astronomers have recognized another of
these shifts of perspective.  They have shown that the stuff that we are
made of accounts for only 4\% of the total content of the universe.
As I will describe, we now know that about 20\% of the energy in  the 
universe takes the form of  a new, weakly interacting form of matter, called
`dark matter'.  The remaining 75\% of the energy of the universe is 
found in the energy content of empty space, `dark energy'.  

Dark energy is the most mysterious of these components.  Its story is 
described in the article of Turner in this volume~\cite{Turner}.
Dark matter, though, is the component that most worries the imaginations
of particle physicists.  What particle is this dark matter made of?  Why 
have we not discovered it at our accelerators?  How does it fit together
with the quarks, leptons, and bosons that we have spent our lives studying?

And, conversely, dark matter is the component that most excites us by the 
possibility of its discovery.  There are strong arguments that the next
generation of particle accelerators, beginning next year with the Large
Hadron Collider (LHC) at CERN, will produce the elementary particles of which
dark matter is made.  How can we recognize them?  How can we prove that these
particles are the ones that are present 
in the cosmos?  And, finally, how can we
use this knowledge to image the dark matter structure of the universe?
I will address all of these questions in this article.

\section{Evidence for Dark Matter}

Although the astronomical picture of dark matter has become much clearer
in the last ten years, the evidence for dark matter goes back to the 
early days of extragalactic astronomy. The evidence for 
dark matter is summarized in a beautiful 1988 review article by 
Virginia Trimble~\cite{Trimble}.  I will describe the most telling 
elements here.

 In 1933, Fritz 
Zwicky measured the mass of the Coma cluster of galaxies, one of the 
nearest clusters of galaxies outside of our local group~\cite{Zwicky}.
  Zwicky's
technique was to measure the relative velocities of the galaxies in this 
cluster from their Doppler shift, use the virial theorem to infer the 
gravitational potential in which these galaxies were moving, and 
compute the mass that must generate the potential.  He found this 
mass to be 400 times the mass of the visible stars in galaxies in the 
cluster.  The observation was soon confirmed by similar measurements of 
the Virgo cluster by Smith~\cite{SSmith}.

We now know that most of the atoms in clusters of galaxies are not seen 
in observations with visible light.  Because these clusters generate
 enormously deep gravitational 
potential wells, it is easy for hydrogen gas from the galaxies to leak 
out and fill the whole volume of the cluster.  These atoms acquire large
velocities and emit X-rays when they collide.  X-ray images show
the clusters as glowing balls of gas.  This does not remove the mystery,
however, The X-ray emitting gas accounts for at most 20\% of the mass of the 
cluster and cannot explain the origin  of the deep potential 
well~\cite{clusters}. For 
this, we must postulate that the clusters are also filled with a new, 
invisible, weakly interacting form of matter.

In the 1970's, astonomers began to systematically measure
the rotational velocity profiles or {\it rotation
curves}, for many galaxies.  One would expect that the mass of a 
galaxy is concentrated in the region where tha stars are visible. Then,
outside this region, Kepler's law would predict that the velocities
should fall off as $1/\sqrt{r}$.  In fact, the velocities are seen to be
constant or even slightly increasing~\cite{Rubin}.  In the 
galaxy NGC 3067, using hydrogen gas lit up by a background quasar, 
Rubin, Thonnard, and Ford 
showed that the rotational velocity profile
maintains its large value at a distance of 
40 kpc (120,000 light-years) from the center of the galaxy, even though the 
visible stars become rare outside of 3 kpc~\cite{TandR}.  From measurements
of the velocities of globular clusters, it was found that the rotation 
curve of our own galaxy is also flat out to distances of 100 kpc from the 
center~\cite{Trimble}.

Detailed measurements of cosmic microwave background, including not only the 
averaged intensity of this background radiation but also its fluctuation
spectrum, give additional information on dark matter.  
  The microwave 
background was emitted at the time of {\it recombination}, when the 
hydrogen filling the 
universe, at a temperature of about 1 eV, converted from an ionized plasma
to a transparent neutral gas. From the Fourier spectrum of fluctuations of
the background radiation, it is possible to measure the dissipation of
this medium.  The most recent measurements from the WMAP satellite require
a medium in which 
only 20\% of the matter is hydrogen gas and 80\% is composed of 
a very weakly interacting species in nonrelativistic motion~\cite{WMAP}.
 These measurements
can be converted to the current fractions of atomic and dark matter in the 
total energy of the universe, $\Omega_i = \rho_i/\rho_{tot}$~\cite{LLreview}
\beq
       \Omega_{at} = 0.042 \pm 0.003  \qquad   \Omega_{DM} = 0.20 \pm .02
\eeq{Omegas}

%%%%%%%%%%%%%%%%%%%%%%%%%%%%%%%%%%%%%%%%%%%%%%%%%%%%%%%%%%%%%%%%%%%%%%%%%
\begin{figure}
\begin{center}
\includegraphics[width=2.9in]{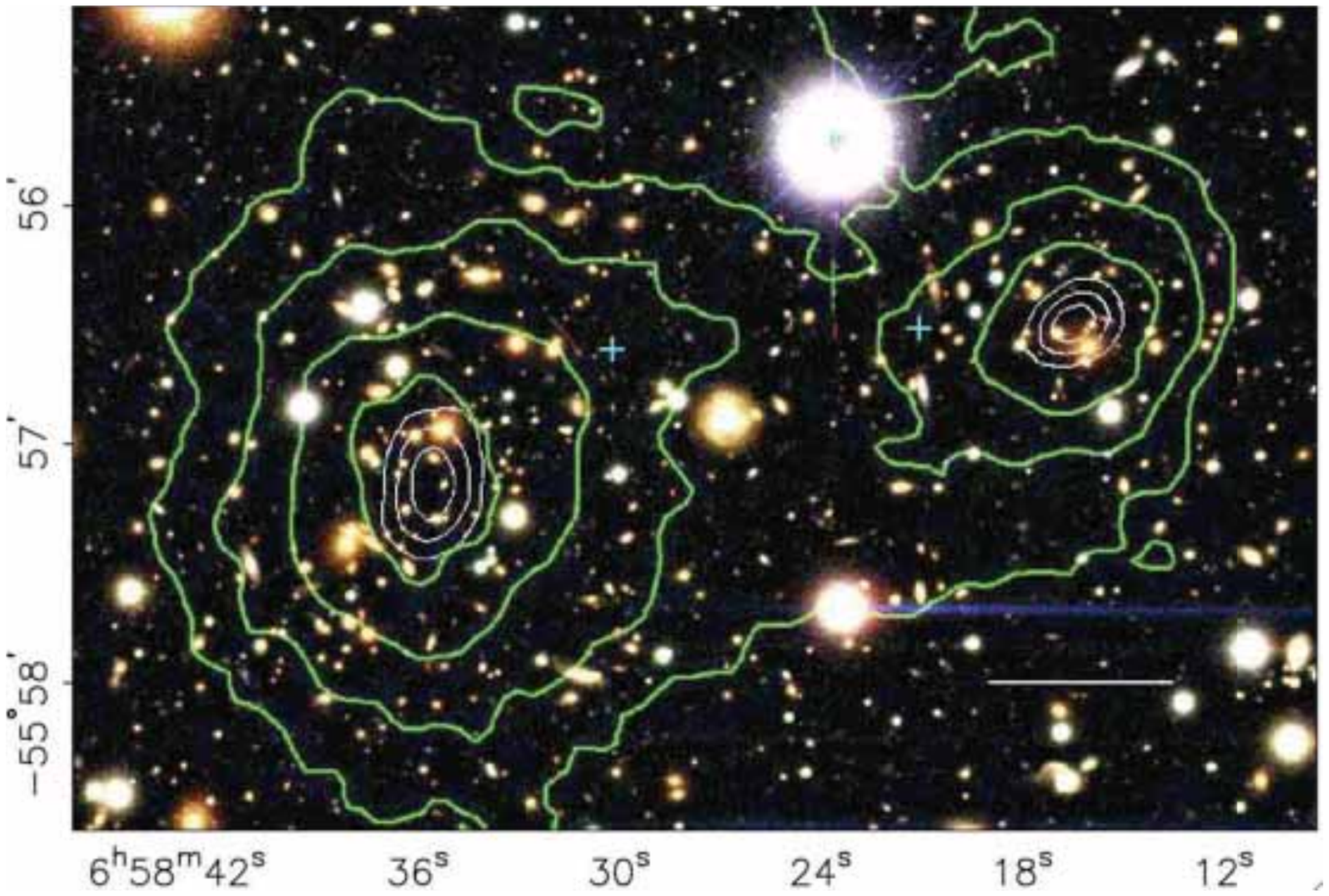}\quad 
\includegraphics[width=2.9in]{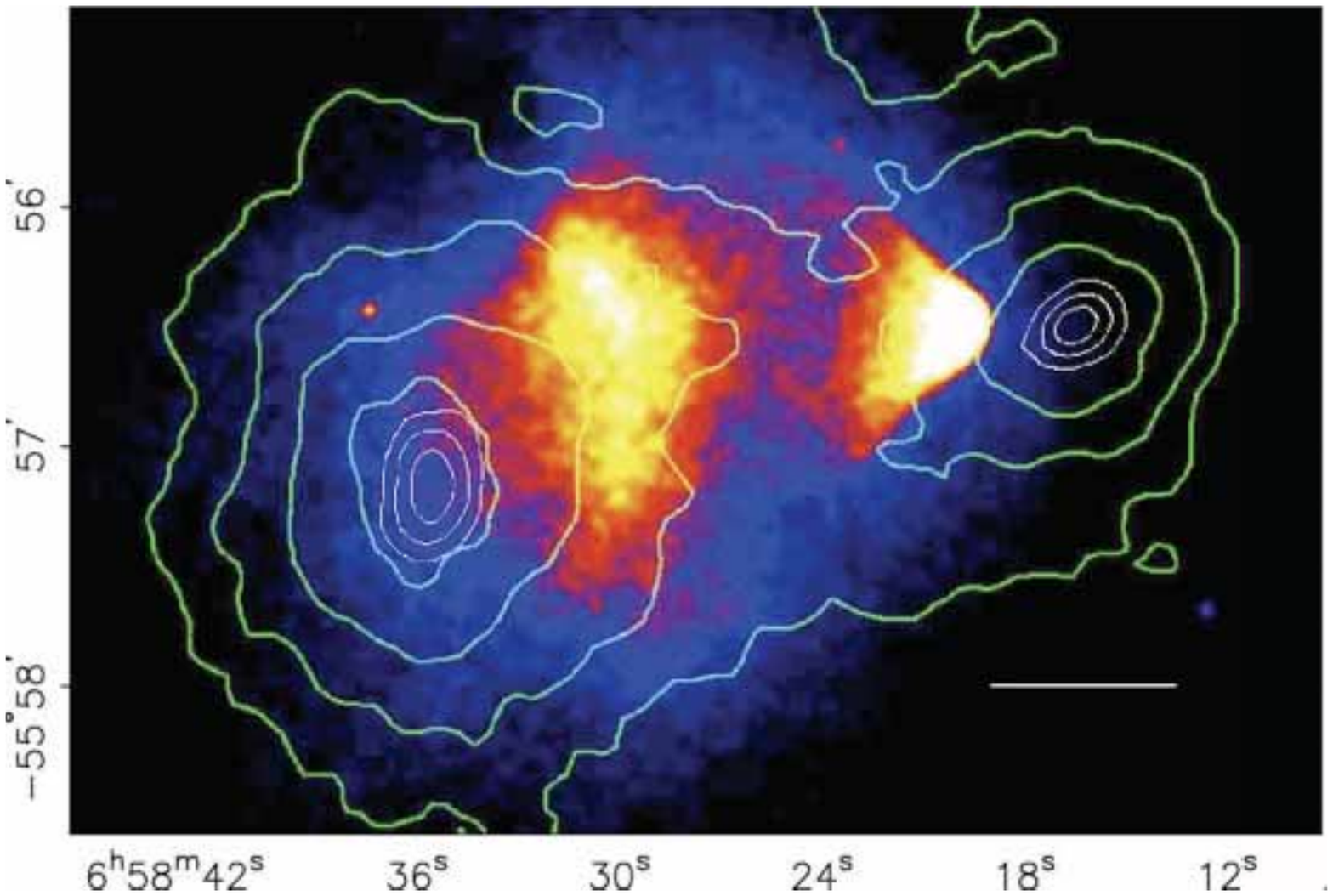} 
\caption{Images of the bullet cluster, 1E0657--558: 
optical image from the Hubble
Space Telescope, X-ray image from the Chandra telescope, and mass density 
contours from gravitational lensing reconstruction~\cite{bullet}.}
\label{fig:bullet}
\end{center}
\end{figure}
%%%%%%%%%%%%%%%%%%%%%%%%%%%%%%%%%%%%%%%%%%%%%%%%%%%%%%%%%%%%%%%%%%%%%%%%%%%

In all of these systems, dark matter is observed only through its gravitational
influence.  One might wonder, then, whether it is possible to explain the 
observations by modifying the law of gravity rather than by introducing a 
new form of matter~\cite{MOND}.  The interpretation in terms of a new form
of matter was recently boosted by the observations shown in 
Fig.~\ref{fig:bullet}.  These picture show three
 images of the galaxy 
cluster 1E0657--558~\cite{bullet}.
  The first is the optical image, showing the galaxies that, 
as we have discussed, make up only a few percent of the mass of the cluster.
The second picture shows
 the X-ray image from the Chandra telescope. This image shows where
the 
bulk of the gas in the cluster is located.  The superimposed
contours show the total density
of the  mass in the cluster, as measured by gravitational
lensing.  It is remarkable that the peaks of the mass distribution
occur where there are very few atoms.  
In this object, which probably arose from a collision of two 
clusters of galaxies, the atomic matter and the dark matter have become 
spatially separated.  The observations cannot be explained by an 
altered law of gravity centered on the atoms.  They require dark matter as a 
new and distinct component.

\section{The WIMP model of Dark Matter}

Thus, dark matter exists.  What is it made of?  In the Standard Model
of particle physics, we know no neutral heavy elementary particles that 
are stable for the lifetime of the universe.  Let us postulate a new 
species of elementary particle to fill this role.  Bahcall called this 
a Weakly Interacting Massive Particle (WIMP).  I would like
to add one more assumption:  Although it is stable, the WIMP
can be produced in pairs (perhaps with its antiparticle), and it was 
produced thermally at an early time when 
the temperature of the universe was  very high.  WIMPs must also annihilate 
in pairs.  I will assume that these processes established a thermal
equilibrium.  

These assumptions lead to an attractive theory of dark matter
whose consequences I will explore in the remainder of this article.
There are other models of dark matter that do not fit into this 
paradigm.  A comprehensive review of dark matter models has recently 
been given by Bertone, Hooper, and Silk~\cite{BHS}. 

Using the WIMP model, we can build a quantitative theory of the density
of dark matter in the universe.  As the universe expanded and cooled, the 
reactions energetic enough to  produce WIMPs became more rare.  But at the
same time, WIMPs had more difficulty finding partners to annihilate.  Thus,
at some temperature $k_B T_f$, they dropped out of equilibrium.  
A small density 
of WIMPs was left over.  At this era, the energy density of the universe 
was dominated by a hot thermal gas of quarks, gluons, leptons, and 
photons, with a total number of degrees of freedom $g_* \approx 80$.  Using
this thermal density to fix the expansion rate of the universe as a function
of time, we can determine the evolution of the WIMP density by integrating 
the Boltzmann equation.  It is convenient to normalize the WIMP density 
to the density of entropy, since in standard cosmology the universe expands
approximately adiabatically.  Then one finds~\cite{TurnerScherrer}
\beq
   \Omega_{DM} =  {s_0\over \rho_{tot}} \left({\pi\over 45 g_*}\right)^{1/2}
      { k_B T_f/mc^2 \over m_\Pl/\hbar^2 \cdot \VEV{\sigma v}}
\eeq{Omegaeq}
where $s_0$ and $\rho_{tot}$ are the current densities of entropy and 
energy in the universe, $m_\Pl$ is the Planck mass, equal to 
$\hbar c/\sqrt{G_N}$, and $\VEV{\sigma v}$ is the thermally averaged
annihilation cross section of WIMP pairs multiplied by their relative
velocity.  In the equation that determines $T_f$, this temperature appears
in a Boltzmann factor $e^{-mc^2/k_B T_f}$, where $m$ is the WIMP mass.  
Taking the 
logarithm, one finds 
$k_B T_f/mc^2 \approx 1/25$ for a wide range of values of the annihilation
cross section.  

With this parameter determined, we know all of the terms in
\leqn{Omegaeq} except for the value of the cross section, and so
we can solve for this factor.  The result is
\beq
       \VEV{ \sigma v} = 1 \ \mbox{pb}
\eeq{sigmavdet}
A particle physicist would recognize this value as the typical size of 
the production cross sections expected for new particles at the LHC.  More
generally, if we assume that the coupling constant in the WIMP interactions
is roughly same size as the dimensionless coupling $\alpha$ that gives the 
strength of weak and electromagnetic interactions,
this cross section results from an
interaction mediated by a particle whose mass is of the order of 100~GeV.

This result is remarkable for two reasons.  First, it allows us to transform
our astrophysical knowledge of the cosmic density of dark matter into a 
prediction of the mass of the dark matter particle.  Second, that prediction
picks out a value of the mass that is very close to the mass scale 
associated with the Higgs boson and the symmetry breaking in the weak 
interactions.  In an earlier article in this volume, Okada has argued that 
we should expect to find new elementary particles at that 
mass scale~\cite{Okada}.
Perhaps these new paricles are in some way responsible for the dark matter.

In fact, explicit models of symmetry breaking in electroweak interactions
often provide a natural setting for dark matter.  Supersymmetry, 
discussed in the article of Yamaguchi in this volume~\cite{Yamaguchi}, 
predicts a new boson for each known fermion
in Nature, and vice versa.  It is natural that the fermionic partner
of the photon is its own antiparticle, so that it is stable but
annihilates in pairs.  This particle is then a perfect candidate for the 
WIMP. Other models of electroweak symmetry breaking also contain new 
neutral weakly-interacting particles that can be made stable by natural
symmetry principles.

\section{Production and Dectection of WIMPs at the LHC}

If the mass of the WIMP should be about 100~GeV, we should be able to 
produce WIMPs if we can build an accelerator that provides elementary 
particle collisions at energies higher than 100~GeV.  However, it is not
so straightforward.  A WIMP, being as weakly interacting as a neutrino, 
passes through a
typical elementary particle detector unseen. It is only from the properties
of the
other particles produced in association with the WIMP that we can 
recognize these events and select them for analysis. Particle physicists have
analyzed in some detail how to do this.  Most of the specific analysis has been
done in models of supersymmetry, so, for concreteness, I will use that
picture here. The general conclusions apply to WIMPs in many other 
models of weak interaction symmetry breaking.

Supersymmetry predicts many new elementary particles in addition to the 
WIMP.  In particular, the gluon of QCD has a fermionic partner, the {\it 
gluino}, and the quarks have bosonic partners, called {\it squarks}.
Gluinos and squarks carry the same conserved quantum number that keeps 
the WIMP stable.  They are expected to be heavier than the WIMP and to decay 
to the WIMP by emitting quarks, leptons, and Standard Model bosons.
Events with squark or gluino pair production, then, will have a characteristic
form.  Many energetic quarks and leptons will be emitted, but also each
event will end with the production of two WIMPs that make no signal
in the detector.  The observable particles in the event will show an 
imbalance of total momentum.  The missing momentum is that carried off
by the WIMPs.

%%%%%%%%%%%%%%%%%%%%%%%%%%%%%%%%%%%%%%%%%%%%%%%%%%%%%%%%%%%%%%%%%%%%%%%%%
\begin{figure}
\begin{center}
\includegraphics[height=2.7in]{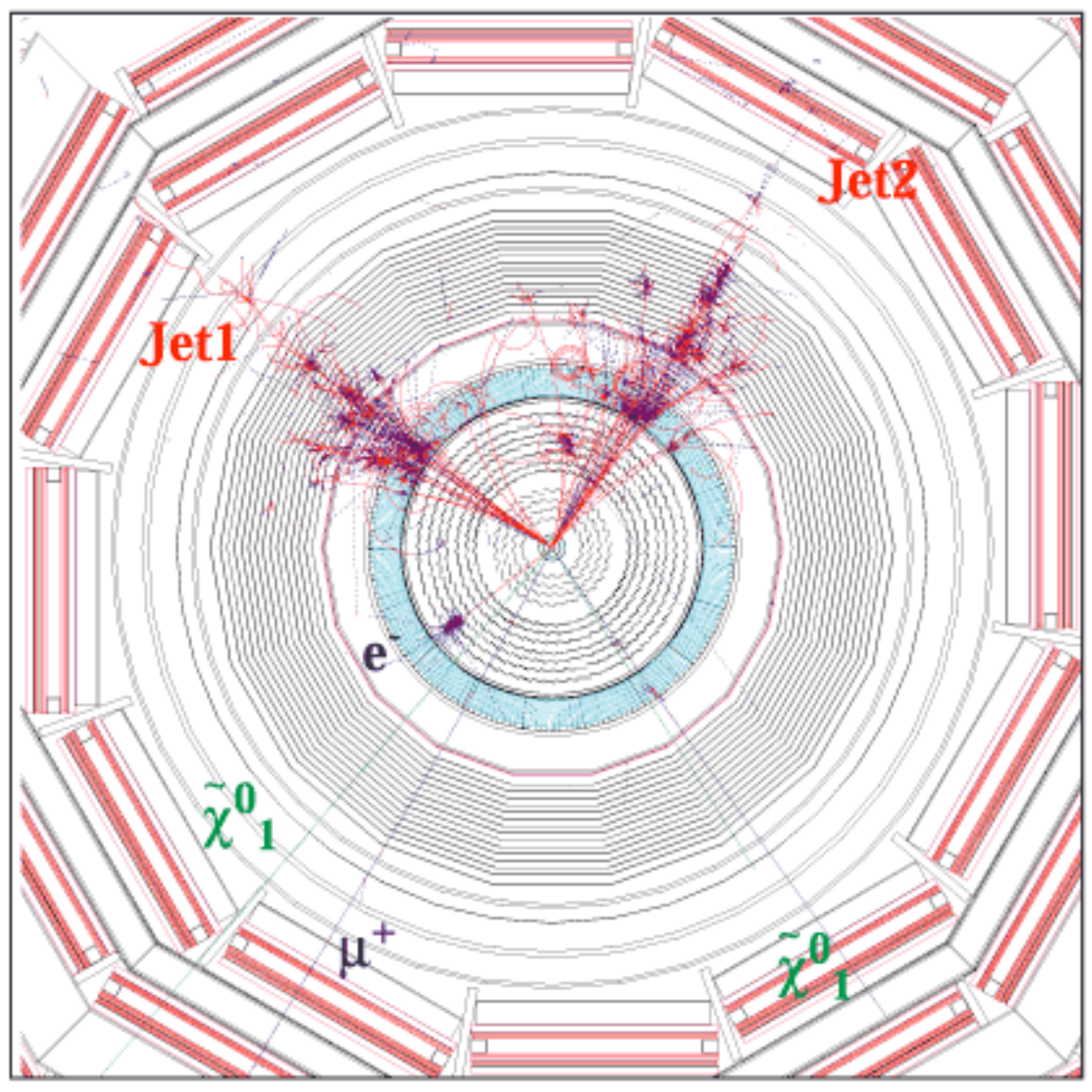}
\caption{Simulated LHC event, with pair production of gluinos and the 
decays of these particles to WIMPs, as would be observed by the CMS 
experiment at the LHC~\cite{CMSevent}.   The apparent imbalance of momentum
transverse to the beam axis is due to the WIMPs (denoted $\widetilde\chi^0_1$ 
in the
figure), which produce no signals in the detector.}
\label{fig:CMSSUSY}
\end{center}
\end{figure}
%%%%%%%%%%%%%%%%%%%%%%%%%%%%%%%%%%%%%%%%%%%%%%%%%%%%%%%%%%%%%%%%%%%%%%%%%%%

We have not yet seen events of this type at currently operating accelerators.
The highest-energy accelerator now operating is the Tevatron collider at 
Fermilab, and the experiments there put lower limits of about 300~GeV on 
the masses of gluinos and squarks~\cite{TevatronSUSY}.  In 2008, however,
the LHC will begin operation with proton-proton collisions at a center of 
mass energy of 14000~GeV.  Not all of this energy is available for production
of supersymmetry particles.  The proton, after all, is a bound state
of quarks and gluons.  Gluinos and squarks are produced in collisions of
individual quarks and gluons, which typically carry 10\% or less of the 
total energy of the proton.  Still, we expect to see collisions with 
total energy above 2000 GeV at a significant rate.  This implies
that squark and gluon pair production, leading to events with WIMPs, can be
seen over almost all of the parameter space of the model.  
Figure~\ref{fig:CMSSUSY} shows a simulation of a  characteristic event of 
this type, as it would be observed by the CMS detector at the 
LHC~\cite{CMSevent}.

\section{Recognizing the Mass of the  WIMP}

The discovery of events at the LHC with apparent unbalanced momentum will
signal that this accelerator is producing weakly interacting massive particles.
However, it would be far from clear that this particle is the same one that 
is the dominant form of matter in the universe.  To demonstrate this, we would
need to correlate properties of the WIMP that we observe at the LHC with
astrophysical observations.  This will probably first be done through 
measurements of the mass of the dark matter particle.  Using detailed
measurements of the kinematics of quarks and leptons in the LHC events, it
is expected that the mass of the WIMP produced there will be measured to 
10\% accuracy~\cite{Paige}.  We then must compare this value with measurements
of the mass of the cosmic WIMP.  To do this, it is necessary to detect the
dark matter in the galaxy, not as a distribution of gravitating mass, but 
as individual particles.

There are two strategies to make this detection.  The first, reviewed by 
Spooner in his article in this volume~\cite{Spooner}, is to place very 
sensitive detectors in ultra-low background environments and look for 
rare events in which a WIMP in our cosmic neighborhood falls to earth 
and scatters from an atomic nucleus in the detector.  The cross section
for this process is expected to have the remarkably small value of 
 1-10 zeptobarns, but in the next few years semiconductor and liquid noble
gas detectors in deep mines are expected to reach this level of sensitivity.
The mean energy deposited in these events depends on the WIMP mass $m$ 
and the target nucleus mass $m_T$ roughly as
\beq
        \VEV{E} = {2 v^2 m_T\over (1 + m_T/m)^2 } \ . 
\eeq{energydep}
Then, for a 100 GeV WIMP, detection of 100 scattering events would lead to a
mass determination at roughly 20\% accuracy~\cite{Lewin,Green}.

The second strategy is to look for WIMP annihilations in our galaxy.
Although the density of WIMPs is sufficiently small that WIMPs cannot 
annihilate frequently enough to affect the overall mass density of the 
universe, WIMPs still should annihilate at a low rate, especially in places
where their density is especially high.   Astrophysicists understand 
the formation of galaxies and larger structures in the universe as 
arising from the clumping of dark matter as a result of its gravitational
attraction.  So our galaxy, and especially the center of the galaxy, 
should be a place with a relatively high density of WIMPs and thus a higher
rate of WIMP annihilations.  In an individual WIMP annihilation, the 
two WIMPs produce two showers of quarks, which are observed mainly as 
pions and photons. The pions and other charged particles are bent by the 
galactic magnetic field. But the photons, energetic gamma rays, fly outward in 
a straight line from the annihilation point.
  A gamma ray telescope can observe 
these photons and measure their energy spectrum.  The spectral shape is 
characteristic, with a sharp cutoff in energy at the mass of the 
WIMP~\cite{Baltzspectrum}.
The galaxy is expected to contain clumps of dark matter that should show up
as spots bright in gamma rays with no counterpart in optical radiation.
These spots should be intense enough to be seen with the gamma ray 
telescope satellite GLAST, and, if the WIMP mass is greater than
several hundred GeV, by new 
ground-based gamma ray telescopes.  Measurement of the endpoint of the
energy spectrum should give a second astrophysical determination of the WIMP
mass to 20\% accuracy. 

If the mass of the WIMP seen at the LHC is the same as the mass from 
astrophysical detection experiments, this will provide strong 
evidence that the LHC is producing the true particle of dark matter.

\section{Predicting the Properties of the WIMP} 

To provide additional evidence on the identity of the WIMP observed at the 
LHC, we would like to assemble enough data about this particle to predict
its pair annihilation cross section.  From \leqn{Omegaeq}, we see 
that knowledge of this cross section from particle physics would give 
a prediction of the cosmic density of 
dark matter.  It will be very interesting to compare that prediction to 
the value of the dark matter density obtained from cosmic microwave 
background measurements.  Agreement of these values would not only confirm
the identity of the WIMP.  It would also verify the standard picture of the 
early universe up to the temperature $T_f$, corresponding to 
a time in the early universe about $10^{-9}$ seconds after the Big Bang.

It is quite a challenge to predict the WIMP pair annihilation cross section.
At the minimum, this requires measuring the masses and couplings of the
heavier particles that are exchanged in the process of WIMP annihilation.
In supersymmetry, WIMP annihilation is often dominated by the exchange of 
the bosonic partners of leptons, which must be identified through their 
decays to leptons and missing momentum.  An alternative mechanism for 
WIMP annihilation is the exchange of the fermionic partners of the weak 
interaction bosons $W$ and $Z$.  These cross sections depend sensitively 
on the mixing angles that determine the exact mass eigenstates of these
particles.   If several different reactions can contribute, the parameters
of each must be measured or bounded.

%%%%%%%%%%%%%%%%%%%%%%%%%%%%%%%%%%%%%%%%%%%%%%%%%%%%%%%%%%%%%%%%%%%%%%%%
\begin{figure}
\begin{center}
\includegraphics[height=2.5in]{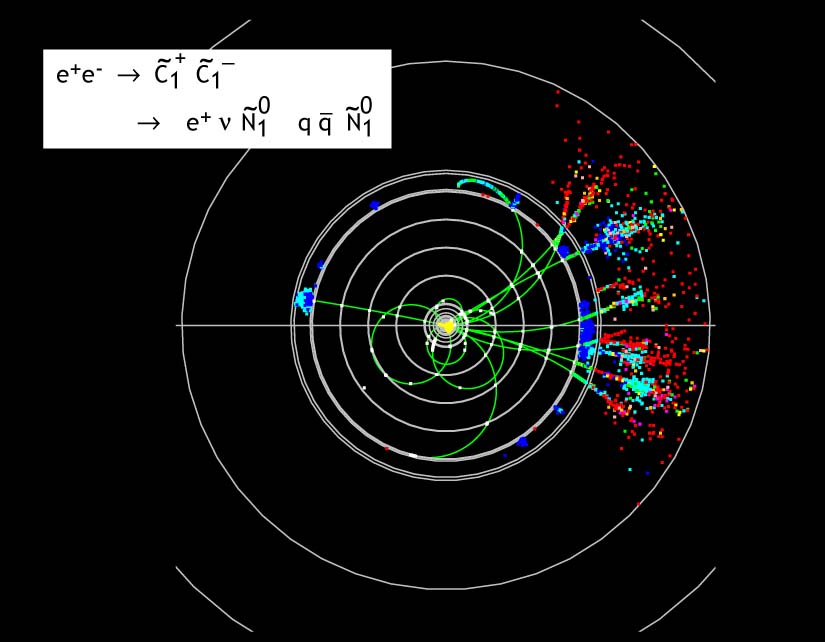}
\caption{Simulated ILC event, with pair production of the supersymmetric
partners of $W$ bosons and subsequent decay to quarks, leptons, and 
WIMPs~\cite{Graf}.  Only the visible products are shown in the figure.}
\label{fig:ILCevent}
\end{center}
\end{figure}
%%%%%%%%%%%%%%%%%%%%%%%%%%%%%%%%%%%%%%%%%%%%%%%%%%%%%%%%%%%%%%%%%%%%%%%%%%%

Detailed studies of this program in a variety of supersymmetry models
 show that it requires more precise knowledge of the parameters of the model
than can be obtained from the LHC.  Fortunately, there is another
technique for producing and studying new elementary particles that is 
capable of achieving higher precision.  Electron-positron annihilation
at high energy can create pairs of the new particles in a controlled 
setting, through reactions that are much simpler than those that we expect
at the LHC.  This process will be studied at the future electron-positron
collider ILC discussed in the contribution of Yamamoto to this 
volume~\cite{Yamamoto}.
A simulated  supersymmetry production event at the ILC is shown in 
Fig.~\ref{fig:ILCevent}.

%%%%%%%%%%%%%%%%%%%%%%%%%%%%%%%%%%%%%%%%%%%%%%%%%%%%%%%%%%%%%%%%%%%%%%%%
\begin{figure}
\begin{center}
\includegraphics[height=2.8in]{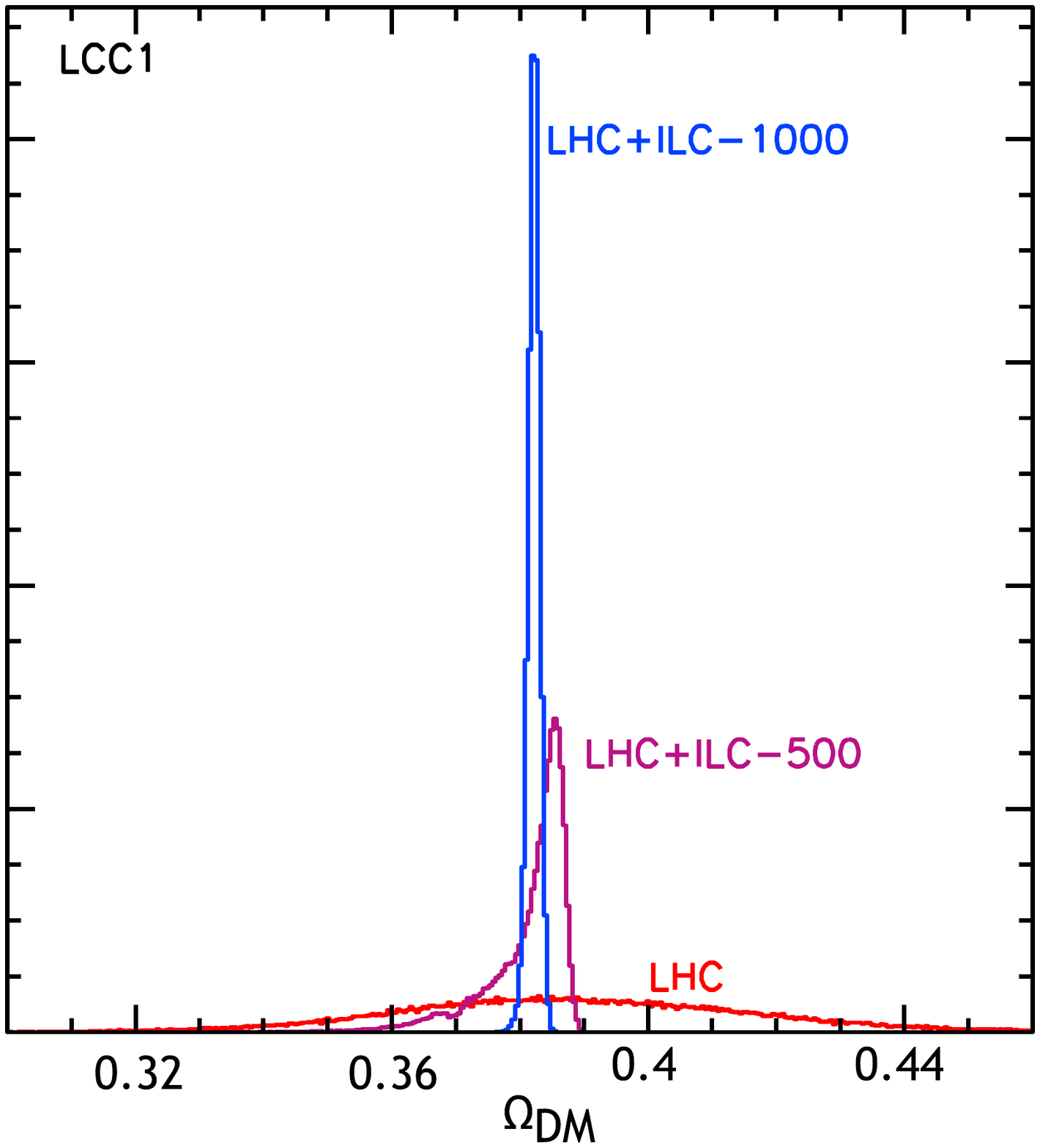}\qquad
\includegraphics[height=2.8in]{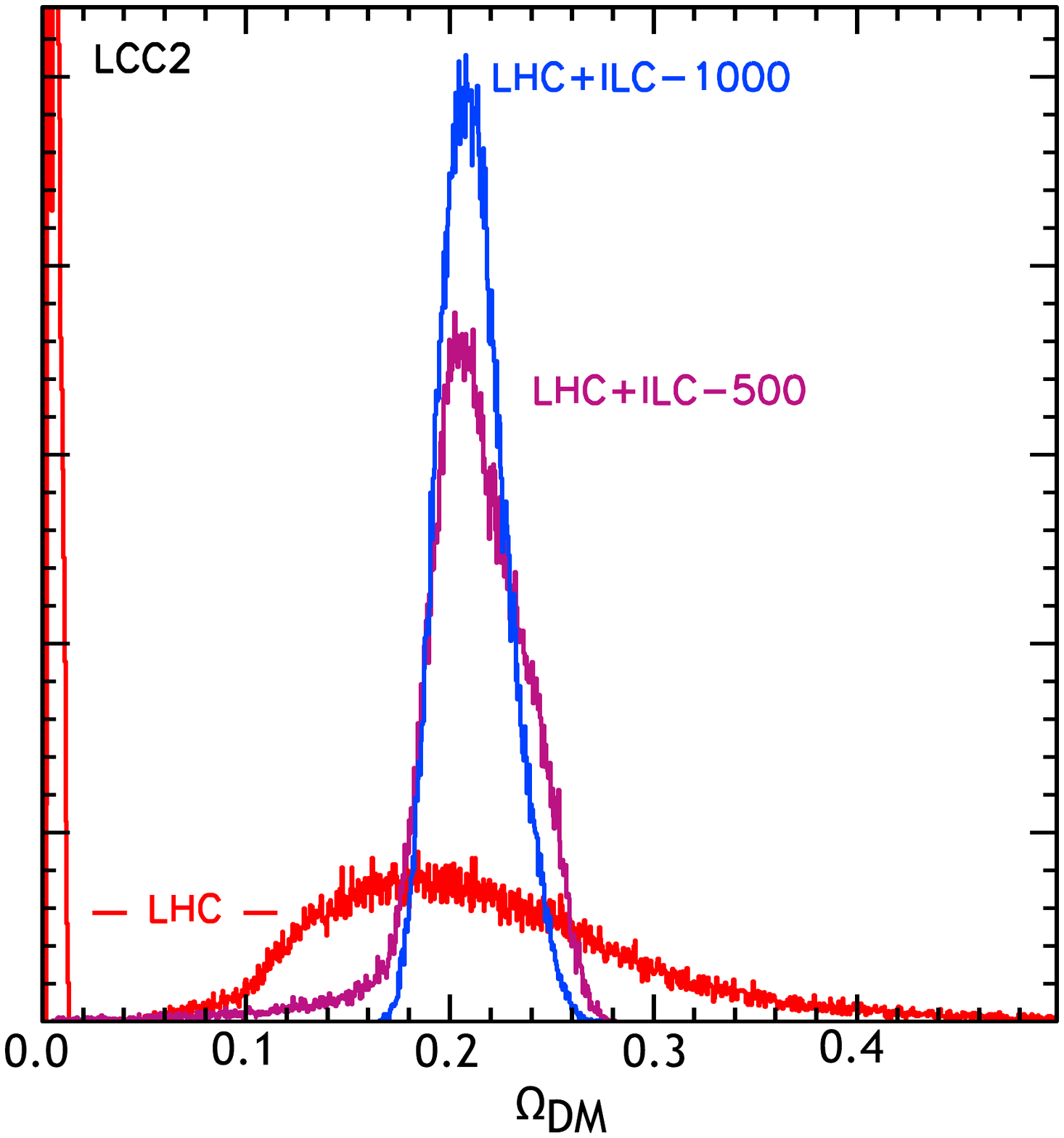}
\caption{Predictions of $\Omega_{DM}$ from 
collider data~\cite{BBPW}.
Each figure was generated by assuming a specific supersymmetry model of
the WIMP, working out the set of measurements that would be made to determine
the spectrum of supersymmetry particles---at the LHC, at the ILC, and at an
upgraded ILC at 1000 GeV in the center of mass---and determining the 
cosmic density of WIMPs from this data.  The figure gives the likeihood 
distribution of the prediction in each model; the accuracy of the collider
measurements determines the  spread in the predictions.}
\label{fig:relic}
\end{center}
\end{figure}
%%%%%%%%%%%%%%%%%%%%%%%%%%%%%%%%%%%%%%%%%%%%%%%%%%%%%%%%%%%%%%%%%%%%%%%%%%%

Once we have measured the masses of supersymmetric particles with high
precision and also measured the cross sections that determine their 
couplings and mixing angles, we will be able to put forward a
prediction of the cosmic dark matter density from particle physics data
that can be compared to 
astrophysical measurements.   Recently, Baltz, Battaglia, Wizansky, and I 
discussed  quantitatively how accurate such microscopic predictions could be.
Starting from a set of supersymmetry models with a variety of different
mechanisms for WIMP annihilation, we analyzed the accuracy of measurements
on supersymmetric particles that could realistically be expected from 
the LHC and from the ILC and derived from these the accuracy of the 
prediction to be expected for the dark matter density~\cite{BBPW}.   
Figure~\ref{fig:relic} shows our results for two
of these models, expressed as the likelihood distribution for $\Omega_{DM}$
predicted from the collider data that would be expected from LHC, from 
ILC measurements at the design energy of 500 GeV, and from an upgraded ILC
running at an energy of 1000 GeV.  Other groups have found similar 
results for first of these models~\cite{Belanger,Nojiri}. 
These predictions will be compared to the cosmic microwave background results
from the next-generation experiments, which should determine $\Omega_{DM}$ 
to the percent level~\cite{Bond}.  It will take some time to collect all of
the data required, but eventually we will have this sharp test of the 
WIMP identity of dark matter.

\section{The WIMP Profile of the Galaxy}

Once we have established the identity and properties of the WIMP, these 
results should feed back into astronomy.  I noted in Section 2 that
it is possible to detect dark matter on cluster scales and to map
its distribution using gravitational lensing.  However, for dark matter
in the galaxy, the gravitational bending of light is not large enough 
effect to provide structure information.  To see where the dark matter is
in our galaxy, we need to map dark matter particles.

\begin{figure}
\begin{center}
\includegraphics[width=4.5in]{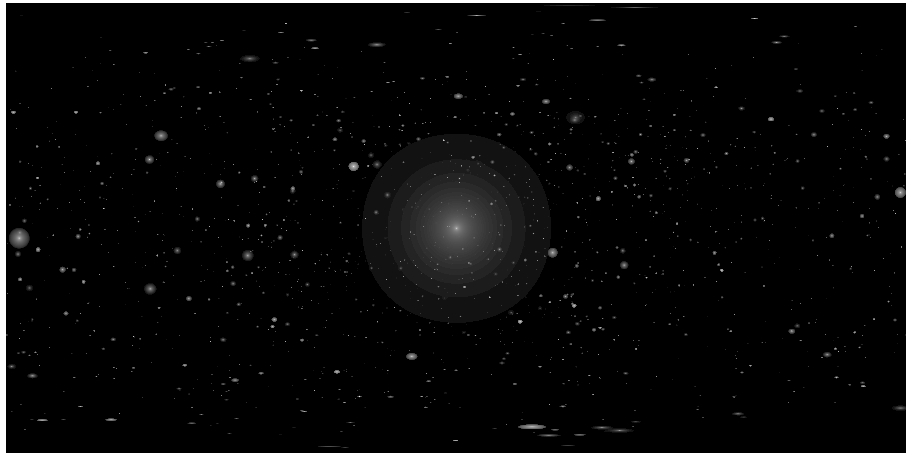}
\caption{Dark matter distribution in a model galaxy, according to the 
simulation of Taylor and Babul~\cite{TaylorBabul}.  This visualization,
done by Baltz~\cite{Baltzpc},  shows a map of the column density 
of dark matter along 
each line of sight.  This quantity gives the brightness with which each
cluster of dark matter shines in annihilation gamma rays.}
\label{fig:TBB}
\end{center}
\end{figure}

The distribution of dark matter in the galaxy is still mysterious, and in 
fact is one of the most controversial questions in astrophysics.  In the 
cold dark matter model of structure formation, a galaxy as large as ours
must be built from the assembly of smaller clusters of dark matter.  
The smaller clusters merge through their gravitational interaction, disrupt
one another tidally, and eventually smooth out to form the halo of the 
galaxy.  The time required for this evolution is on the order of the 
current age of the universe.  Thus, most cold dark matter theories predict
that the halo of the galaxy is inhomogeneous.  A model of the density 
distribution of dark matter in a model galaxy, based on the clustering 
model of Taylor and Babul \cite{TaylorBabul} is shown in Fig.~\ref{fig:TBB}.
An especially large 
clustering of dark matter should occur at the center of the galaxy.
Some models predict caustics with large, almost singular dark matter 
densities; other models predict smoothing of the dark matter below some
scale.  Understanding the true situation will bring us closer to understanding
how our galaxy and the others in the universe were born and 
evolved~\cite{Primack}.

The determination of the properties of the dark matter particle will give us 
the information that is needed to predict the interaction rates of dark 
matter particles with ordinary matter and with one another.  This in turn
will allow us to interpret detection signals in 
terms of the absolute density of dark matter both here and elsewhere in the
galaxy.  By dividing the underground detection rate for dark matter by 
the interaction cross section determined from collider data, we will be 
able to measure the absolute flux of dark matter at our position in the 
galaxy.  By measuring the luminosity of clumps of dark matter in the 
galaxy and dividing by the dark matter annihilation cross section 
determined from collider data, we will be able to map at least the largest
clumps of dark matter in terms of their absolute density.

%%%%%%%%%%%%%%%%%%%%%%%%%%%%%%%%%%%%%%%%%%%%%%%%%%%%%%%%%%%%%%%%%%%%%%%%
%%%%%%%%%%%%%%%%%%%%%%%%%%%%%%%%%%%%%%%%%%%%%%%%%%%%%%%%%%%%%%%%%%%%%%%%%%%

\section{Conclusions}

Today, dark matter is one of the great mysteries of physics and astronomy.
But I have argued in this article that the time is approaching for its
solution.  I have motivated the idea that dark matter is composed of a new
elementary particle, the WIMP, whose mass is about 100 GeV.  If this is true,
then over the next five years
we should produce the WIMP at the LHC, and we should see signals of 
astrophysical WIMPs in several different detection experiments.  This will 
set in motion a campaign to determine the properties of dark matter by 
measurements both in high-energy collider experiments and through mapping
of astrophysical signals.  Over the next fifteen years, we will learn the 
story of this major constituent of the universe, its identity, its properties,
and its role in our cosmic origin.

\Acknowledgements

I am grateful to Marco Battaglia, Ted Baltz, Jonathan Feng, Mark Trodden,
Tommer Wizansky, and many others whose insights were essential
in creating the picture explained here.  I thank David Griffiths for 
instructive comments on the manuscript. This work was supported by the
U S Department of Energy under contract  DE--AC02--76SF00515.

\end{document}